# An Edge-Extrusion Approach to Generate Extruded Miura-Ori and Its Double Tiling Origami Patterns

K. Suto, A. Adachi, T. Tachi, Y. Yamaguchi

**Abstract:** This paper proposes a family of origami tessellations called extruded Miura-Ori, whose folded state lies between two parallel planes with some faces on the planes, potentially useful for folded core materials because of face bonding. An extruded Miura-Ori is obtained by cutting Miura-Ori apart along the edges and face diagonals before inserting the extrusion of the section edges. We compute the extrusion direction to obtain a valid extruded Miura-Ori. The extruded Miura-Ori usually has three valid states. We analyse the third state (final folded state) in depth to show that a continuous family of parameters can produce origami tessellations that can completely tile the top and bottom planes.

## 1  Introduction

A sandwich panel is a lightweight and high-stiffness structure made of a low-density core and thin layers (often called skin-layers) bonded to both sides. We focus on sandwich panels whose cores are made of sheet materials. For example, in our daily lives, corrugated cardboards are widely used in the manufacturing of shipping containers and corrugated boxes. They are resistant to bending in a specific direction but weak when bending in the perpendicular directional because of an anisotropic core structure. In situations where high performance is required, honeycomb core sandwich panels are used, which consist of skin-layers and a core material filled by closed hexagonal prism cells similar to honeycombs. In contrast to corrugated cardboards, honeycomb core sandwich panels exhibit stiffness in all directions but cannot pass fluids as they are composed of closed cells.

  An origami core sandwich panel is made of an origami tessellation folded into a 3D panel with some thickness as a core material and two flat planes, called *skin-planes*, affixed to the two sides of the panel. The origami core sandwich panel has attracted considerable attention from engineers for two reasons: the core material can be manufactured by folding a piece of planar material, and an origami core sandwich panel made using Miura-Ori can pass fluids because it consists of open cells [Miura 1970, Klett et al. 2010]. The origami core sandwich panel potentially offers higher performance than the honeycomb core sandwich panel under certain situations often required for the purpose of temperature control. However, when an origami core such as Miura-Ori is bonded to skin-planes, *bonding regions*—where the parts of the origami tessellation lying on skin-planes are merely edges with zero areas—potentially make the bonding between the core and the skin weak (Figure 1, upper left). We are interested in exploring the design of origami tessellations that have faces lying on skin-planes. We call such origami tessellations *face-bondable origami tessellations*, defined in Section 2.1. This class of origami tessellations is potentially useful for designing new lightweight folded-core structures (Figure 1).

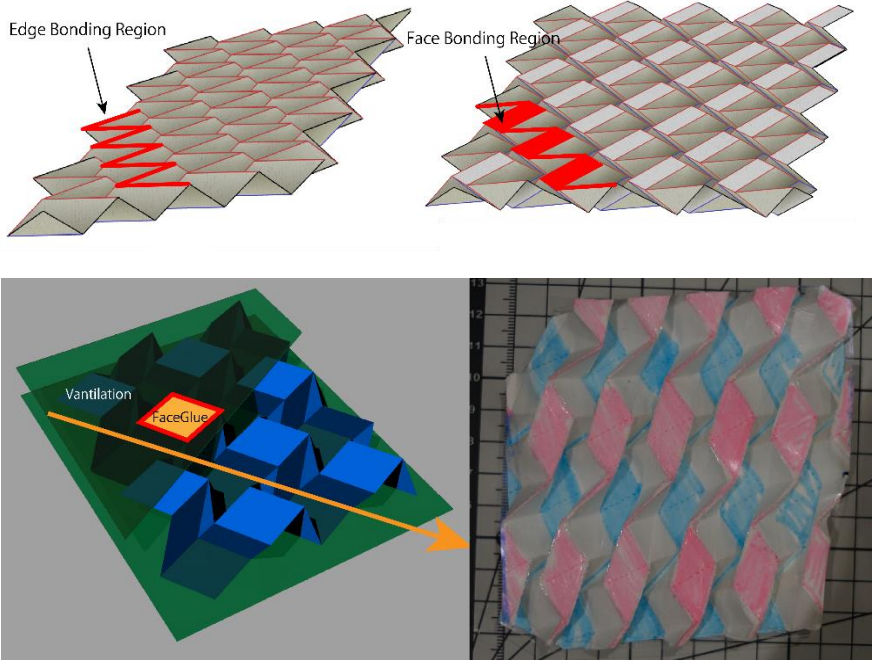

**Figure 1:** *Upper left: A Miura-Ori has zero-area bonding regions. Upper right: A face-bondable origami tessellation obtained by our approach. Lower left: A face-bondable origami sandwich core panel. Lower right: A fabricated model made of a sheet of paper and lamination sheets. The pink and blue faces lie on the top and bottom skin-planes, respectively.*

This paper presents a family of face-bondable origami tessellations called *extruded Miura-Ori.* As the name suggests, an extruded Miura-Ori is obtained by zig-zag cutting a Miura-Ori apart along the edges and the diagonals of the faces and then inserting the extrusion of the section edges. We illustrate how to generate the extruded Miura-Ori and then prove that the method provides a face-bondable origami tessellation in Section 2. In Section 3, we further analyse the folded states and the folding motion of the obtained extruded Miura-Ori to identify the existence of a final folded state, which is also a face-bondable origami tessellation. We demonstrate a parametric design of the extruded Miura-Ori in the final folded state to show some special cases where the top and bottom parallelograms tile the skin-planes.

## 2   How to Generate the Extruded Miura-Ori

### 2.1   Edge-extrusion approach for face-bondable origami tessellations

First, we will state some basic definitions and terminology required to explain our approach. A *face-bondable origami tessellation* is an origami tessellation with the following properties:
- The folded state exists between two parallel planes (we call them *skin-planes*)
- Some of the faces lie on the top skin-plane and some other faces lie on the bottom plane.

Before demonstrating the approach for generating extruded Miura-Ori, we first define a Miura-Ori (Figure 2, left) as an origami tessellation whose structure consists of one parallelogram shape and its mirror. We call this parallelogram the *Miura-parallelogram*. We define an *oblique cutting line* as a polyline composed by alternately connecting the oblique edges (a set of parallel edges that are not edges constituting the line of mirror symmetry) and the diagonals of the parallelograms.

Now, an extruded Miura-Ori is obtained by cutting a Miura-Ori apart along oblique cutting lines and then inserting the parallelogram strips obtained by extruding the section edges along a particular direction offset by a positive distance. We call this direction the *extrusion direction*, the distance the *extrusion depth*, and the obtained parallelogram strip the *extruded strip*. An extruded strip is composed of two types of alternating parallelograms. Parallelograms of one type lie on either of the two skin-planes; we call these parallelograms *skin-parallelograms*.

The extrusion direction must be determined such that the generated polyhedral surface, namely a skin parallelogram, satisfies the following conditions in order that the generated shape is a face-bondable origami tessellation.

1. Developability: The vertices created by the extrusion procedure are developable

2. Face-bondability: Two sets of parallelograms generated in the extrusion direction lie on the top and bottom skin-planes; in particular, the extrusion vector must be parallel to the skin-planes.

In subsection 2.3, we prove there is a unique direction for the folded state of a Miura-Ori that satisfies conditions 1 and 2.

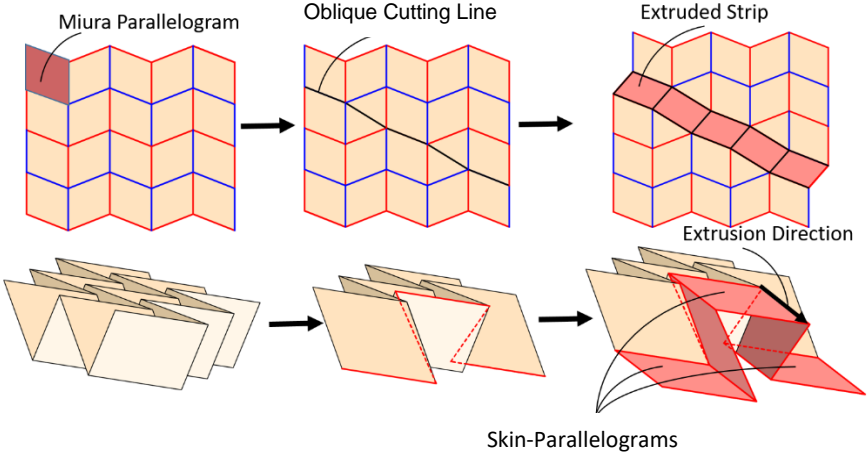

**Figure 2:** *An outline of the edge extrusion approach with notations.*

## 2.2 Developability condition of extruded shapes

Before the Miura-Ori edge-extrusion operation, we consider an extrusion around a general single vertex. Refer to Figure 3 for notations.

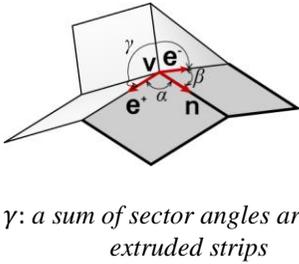

**v**: a *vertex to be extruded*
$\mathbf{e}^+, \mathbf{e}^-$: *each unit vector along cut edge*
**n**: an *extrusion direction unit vector*
$\alpha, \beta$: *extruded strip angles*
$\gamma$: *a sum of sector angles around* **v** *except* $\alpha, \beta$ *extruded strips*

**Figure 3:** *Notations of extrusion around a single vertex.*

The developability condition at **v** is given as:

$$2\pi - (\alpha + \beta + \gamma) = 0$$

which can be represented using vector dot products [Kilian, 2008]:

$$((\cos\gamma)^2 - 1)\|\mathbf{n}\|^2 + (\mathbf{e}^+ \cdot \mathbf{n})^2 + (\mathbf{e}^- \cdot \mathbf{n})^2 - 2(\cos\gamma)(\mathbf{e}^+ \cdot \mathbf{n})(\mathbf{e}^- \cdot \mathbf{n}) = 0. \tag{1}$$

Refer to Figure 4. This equation states that the extrusion direction **n** lies on a spherical ellipse on the unit sphere. In spherical trigonometry, a *spherical ellipse* is defined by two foci ($C$ and $C'$) and a string with length $L$ such that the locus $P$ satisfies $arc\ length(CP) + arc\ length(C'P) = L$ [Maeda 2005]. $\mathbf{e}^+, \mathbf{e}^-, \mathbf{n}, \alpha, \beta$, and $2\pi - \gamma$ correspond to $C, C', P, arc\ length(CP), arc\ length(C'P)$, and $L$, respectively.

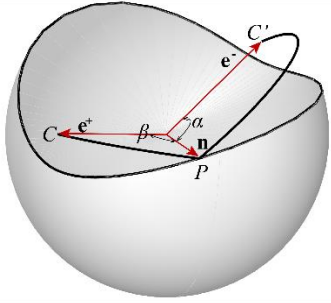

**Figure 4:** *The extrusion vector $\mathbf{n}$ lies on a spherical ellipse.*

Now, we consider an extrusion operation along an oblique cutting line. As there are multiple edges to be extruded, the extrusion direction is given by one of the intersections of spherical ellipses on the unit sphere. So, in general, we require that the number of classes of vertices to extrude under the symmetry be no more than two. The oblique cutting line of the Miura-Ori consists of two classes of vertices under symmetry, as we prove in Section 2.3. Also note that the conditions for the split vertex on the other side are in general the same as the original vertex before the split is developable: flipping $\mathbf{n}$ and $-\mathbf{n}$ $\cos\gamma = \cos(2\pi - \gamma)$ in Equation (1), so the extruded strip is inserted between the Miura-Ori.

## 2.3 Uniqueness of the extrusion direction of a Miura-Ori

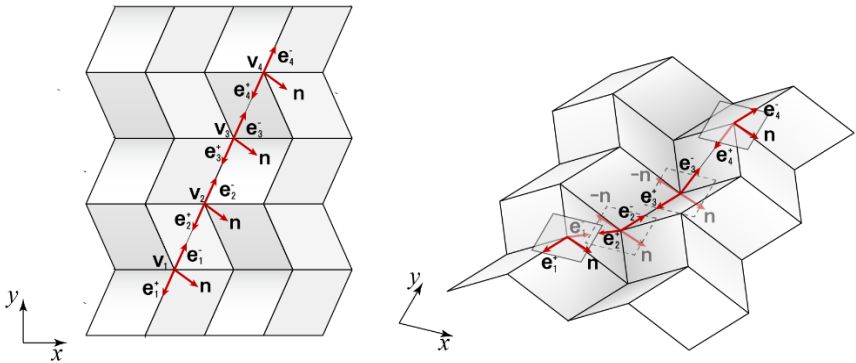

**Figure 5:** *The geometry of a Miura-Ori and four classes of vertices with notations for extrusion direction unit vectors and cut edge unit vectors.*

To compute the extrusion direction of a Miura-Ori, we define the following parameters (refer to Figure 5): a Miura-Ori consists of only a single type of parallelogram. Along the oblique cutting line, four classes of vertices $\mathbf{v}_i$ ($i$ =1, 2, 3, 4) appear. For each vertex $\mathbf{v}_i$, cut edge unit vectors $\mathbf{e}_i^+$ and $\mathbf{e}_i^-$ are defined. Four vertices share a common extrusion unit vector $\mathbf{n}$ (Figure 5 left). It is sufficient to consider two types of vertices because of symmetries. Conditions for $\mathbf{v}_i$ with vectors $\mathbf{n}$, $\mathbf{e}_i^+$, $\mathbf{e}_i^-$ ($i = 1, 4$) are equivalent to conditions for $\mathbf{v}_j$ with vectors $-\mathbf{n}$, $\mathbf{e}_j^-$, $\mathbf{e}_j^+$ ($j = 2, 3$) respectively because of rotational symmetry about the Y axis (Figure 5 right). Note that Equation (1) has a flip symmetry of $\mathbf{n}$ and $-\mathbf{n}$. Thus, we can identify $\mathbf{v}_i$ ($i = 1, 4$) with $\mathbf{v}_j$ ($i = 2, 3$).

Conditions for $\mathbf{v}_i$ ($i = 1, 3$) lead to the extrusion direction $\mathbf{n}$ being parallel to the XY plane because of mirror symmetry. Conditions for $\mathbf{v}_1$ are equivalent to the XY-plane mirror of conditions for $\mathbf{v}_3$. This makes each spherical ellipse produced from these two vertices symmetric across the XY plane. Thus, a solution exists on the XY plane that satisfies the face-bondability condition as well. Consequently, the solution can be given by the intersection between the spherical ellipse defined by $\mathbf{v}_1$ and the XY plane, so it is sufficient to consider the condition for $\mathbf{v}_1$. Therefore, we will henceforth omit the index. We define the parameters of a Miura-Ori in Figure 7 and denote $\mathbf{e}^\pm = [x^\pm \quad y^\pm \quad z^\pm]^T$, $z^+ = 0$.

**Theorem.** *Let* $\mathbf{X} = \mathbf{e}^- - \mathbf{R}_{xy}(-\gamma)\mathbf{e}^+$, *then the extrusion direction vector* $\mathbf{n}$ *that satisfies the developability and face-bondability conditions are uniquely determined, and*

$$\mathbf{n} = \mathbf{X}^\perp := \mathbf{R}_{xy}\left(\frac{\pi}{2}\right)\mathbf{X} \qquad (2)$$

*where* $\mathbf{R}_{xy}(\theta)$ *is a rotational matrix around the Z axis with angle* $\theta$.

*Proof.* **Necessity:** We construct the coordinate system of an extrusion operation as in Figure 5 and 6. By the face-bondability property, an extrusion direction vector $\mathbf{n}$ must be parallel to the XY plane. Also, $\mathbf{n}$ must exist in the direction obtained by rotating the cut edge unit vector $\mathbf{e}^+$ counter-clockwise by an angle $\in (0, \pi)$ in the XY plane viewed from the +z direction; otherwise, obtained faces will be flipped. If the faces of the extruded strip are flipped, self-intersections around the extrusion vertices will result.

Now, consider developability. Because the cut edge unit vector $\mathbf{e}^-$ is connected to $\mathbf{n}$ through a single facet $\mathbf{e}_{\text{flat}}^-$, the developed state of $\mathbf{e}^-$ must be obtained by rotating vector $\mathbf{e}^-$ around $\mathbf{n}$. Thus, $\mathbf{e}^-$ and $\mathbf{e}_{\text{flat}}^-$ are on the same cone with axis $\mathbf{n}$, and then $\mathbf{X} = \mathbf{e}^- - \mathbf{R}_{xy}(-\gamma)\mathbf{e}^+$ and $\mathbf{n}$ are orthogonal. Note that $\mathbf{e}^-_{\text{flat}}$ is also obtained by rotating vector $\mathbf{e}^+$ along the XY plane by $\gamma$. If $\mathbf{n} = \mathbf{R}_{xy}\left(-\frac{\pi}{2}\right)\mathbf{X}$, this rotation causes the directions of the faces of the extruded strip to be flipped. Therefore, $\mathbf{n} = \mathbf{R}_{xy}\left(\frac{\pi}{2}\right)\mathbf{X}$.

**Sufficiency:** (2) satisfies developability. Also, face-bondability is satisfied because $\mathbf{n}$ is parallel to the XY plane and is in the counter-clockwise direction with respect to $\mathbf{e}^+$. □

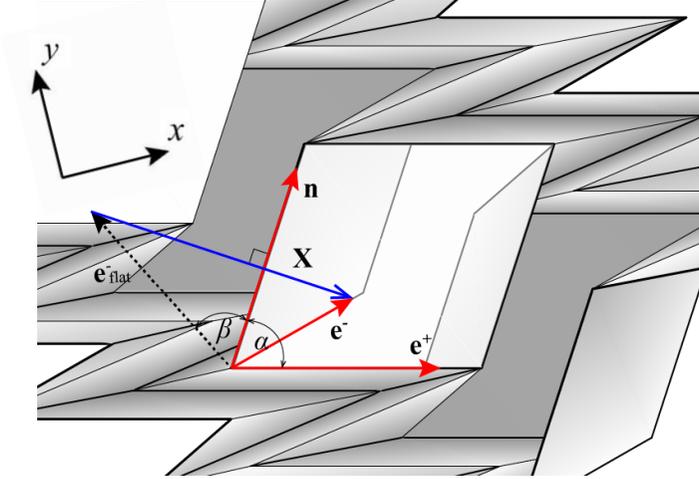

**Figure 6:** *Construction of the extrusion unit vector $\mathbf{n}$.*

## 2.4 Expressions using parameters of Miura-Ori

Because a Miura-Ori has a 1-DOF rigid folding motion, the folded state can be given by the parameters of the shape of the parallelogram $w$, $l$, $and$ $\theta$ ($w$ is the length of the parallelogram edge lying on the line of mirror symmetry, $l$ is the length of the oblique crease of the parallelogram of the Miura-Ori, and $\theta$ is the sector angle inside the parallelogram of the Miura-Ori) and the fold angle $\rho$ of the crease lying on the line of mirror symmetry. We can get the following equations using $h\cos\frac{\rho}{2} = l\sin\xi$ and $\frac{h\sin\frac{\rho}{2}}{\tan\zeta} = l\cos\xi\cos\zeta$ from Figure 7:

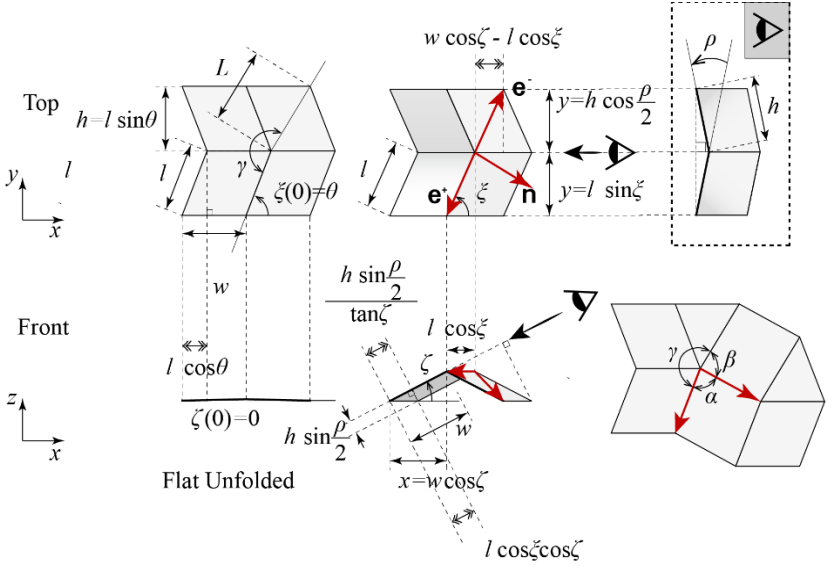

**Figure 7:** *The geometry of Miura-Ori using parameters l, w, θ, and ρ.*

$$\begin{bmatrix} \cos \xi \\ \sin \xi \end{bmatrix} = \begin{bmatrix} \sin \theta \cos \frac{\rho}{2} \\ \sqrt{\left(\sin \theta \sin \frac{\rho}{2}\right)^2 + \cos^2 \theta} \end{bmatrix}, \begin{bmatrix} \cos \zeta \\ \sin \zeta \end{bmatrix} = \begin{bmatrix} \frac{\cos \theta}{\sin \xi} \\ \frac{\sin \theta \sin \frac{\rho}{2}}{\sin \xi} \end{bmatrix}.$$

Using these equations, we can also derive equations for $x^+, y^+, x^-, y^-, \cos \gamma,$ and $\sin \gamma$.

$$\begin{bmatrix} x^+ \\ y^+ \end{bmatrix} = \begin{bmatrix} -\cos \xi \\ -\sin \xi \end{bmatrix} \text{ and } \begin{bmatrix} x^- \\ y^- \end{bmatrix} = \begin{bmatrix} \frac{w \cos \zeta - l \cos \xi}{L} \\ -\frac{l \sin \xi}{L} \end{bmatrix}$$

$$\begin{bmatrix} \cos \gamma \\ \sin \gamma \end{bmatrix} = \begin{bmatrix} \frac{l \cos 2\theta - w \cos \theta}{L} \\ \frac{(\sin \theta)(w - 2 l \cos \theta)}{L} \end{bmatrix},$$

where $L = \sqrt{w^2 + l^2 - 2wl \cos \theta}$.

Thus, we can represent the extrusion direction vector **n** as a function of $w, l, \theta,$ and $\rho$ by a series of substitutions.

## 3  The Final Folded State and Double-Tiling Origami

In this section, we further analyse the folded states and the folding motion of the obtained extruded Miura-Ori. Our observations show that there are three folded states of generic extruded Miura-Ori that makes the skin-parallelograms lie on skin-planes (Section 3.1). These folded states are the developed state, the state of geometric construction (the state right after the generation described in Section 2), and the state where adjacent skin-parallelograms touch. We call the last state the *final folded state* as the contact between parallelograms prevents further folding motion (Figure 8).

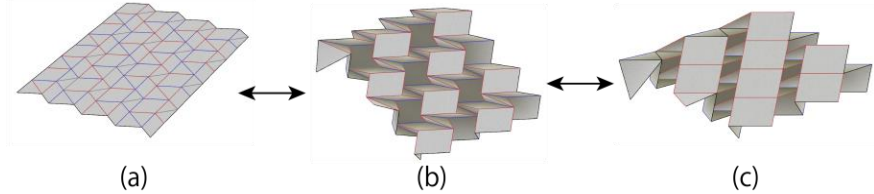

**Figure 8:** *Three folded states of extruded Miura-Ori whose skin-parallelograms lie on skin-planes: (a) the developed state, (b) the state of geometric construction, and (c) the final folded state.*

In Section 3.2, we provide an interpretation of the final folded state, which is still an extruded Miura-Ori but obtained using a different folding mode.

In Section 3.3, we show that under some parametric conditions, the extruded Miura-Ori can have a special final folded state in which the skin parallelograms tile the two planes. Furthermore, the tiling pattern can be further classified into three types, namely *prism type*, *hip-roof type*, and *pyramid type*.

### 3.1 Observation of simulated folding motion

We have simulated the folding motion of extruded Miura-Ori using freeform origami [Tachi. 2010]. In fact, the extruded Miura-Ori seems to not be rigidly foldable. However, when the skin-parallelograms are triangulated, it starts to continuously fold and unfold. Let $\varphi$ be the fold angle of the oblique crease line of a Miura-Ori and $\sigma$ be the fold angle of the diagonal line of triangulated skin-parallelograms. Figure 9 shows the folding motion of extruded Miura-Ori generated with parameters $(l, w, \theta, \rho, d) = (14.260, 10.000, 0.223\pi, 0.756\pi, 14.294)$. We can see that there are three states with $\sigma = 0$.

This suggests that in generic cases, the original extruded Miura-Ori without triangulation has three folded states. These three correspond to the completely unfolded state $\varphi = 0$, the state in which the extruded Miura-Ori is generated $\varphi = \varphi_{\text{flat}}$, and the final state $\varphi = \pi$. In the final state, we can observe that skin-parallelograms in one column collide with each other and share their edges, and they lie on exactly two planes. We conjecture that there is a range of parameters for extruded Miura-Ori that allow the final state. The final folded state will provide another class of face-bondable origami tessellations.

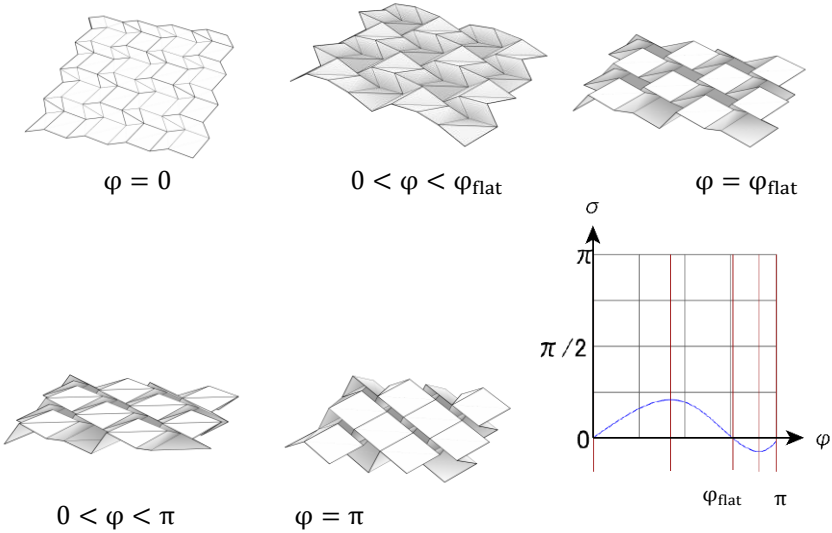

**Figure 9:** *The folding motion of an extruded Miura-Ori and the relationship between its folding angles.*

### 3.2 An interpretation of the final folded state

The final folded state of an extruded Miura-Ori is a face-bondable origami tessellation. Moreover, it can be viewed as an edge extrusion of some origami tessellation or, more precisely, a different folded state of the same Miura-Ori pattern. We have only focused on the "ordinary" folding mode, as shown in Figure 5; however, Miura-Ori have another folding motion exhibiting periodicity. It may be constructed by first zig-zag folding straight lines along the line of symmetry up to 180° and then folding alternating accordion-like strips.

The extrusion operation is the same as the ordinary mode of extruded Miura-Ori we saw in Section 2. Figure 10 shows the extrusion process. It is interesting to observe that self-intersection happens for a small extrusion depth, while for sufficiently larger extrusion depths, we obtain a valid folded state without self-intersections.

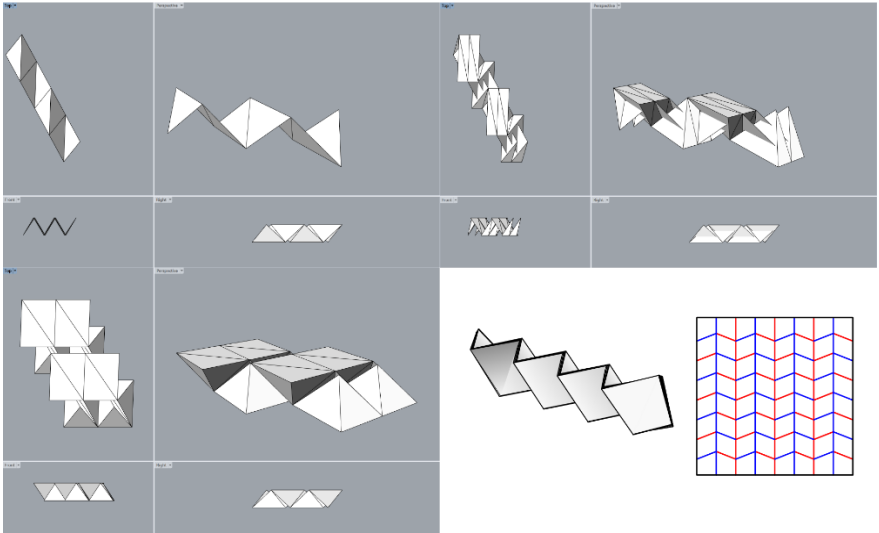

*Figure 10:* An edge-extrusion process applied to a Miura-Ori folded in another mode.

### 3.3 The conditions and classification of double-tiling origami

The final folded state of an extruded Miura-Ori usually has columns of connected skin-parallelograms, where each column is separated by some finite gap (Figure 11, left). We can adjust Miura-Ori parameters to fill the gaps between blocked parallelogram strips. The gap distance $g$ can also be represented as function of $(l, w, \theta, \rho)$. The final folded state of an extruded Miura-Ori with gap distance $g = 0$ has face parallelograms completely tiling the planes. We call a subset of face-bondable origami tessellations whose faces on the skin-planes tile the planes a *double-tiling origami* tessellation.

Within the family of final folded states of extruded Miura-Ori that make up double-tiling origami, we can still change parameters to tweak the phase of the aligning columns of parallelograms. The shift distance of the phase, denoted by $s$, changes the inner structure of the space filling (see Figure 11, right and Figure 12). Table 13 shows how the change in $s$ from 0 to $d$ affects polyhedral filling between skin-planes, where $d$ is the depth of extrusion. In an extreme case of $s = d$, we obtain the dual-tiling origami proposed in previous research [Adachi et al. 2017] (in fact, our extruded Miura-Ori is inspired by the relationship between Miura-Ori and dual-tiling origami, as shown in the Appendix). Refer to Figure 14 for subset relationships between the different classes of origami tessellations. We used *Grasshopper* to define a parametric design of extruded Miura-Ori in its state of generation and at the final folded state and to numerically solve the parameters of special cases of $s = 0$ and $g = 0$. Figure 17 shows the table of extruded Miura-Ori with critical parameters.

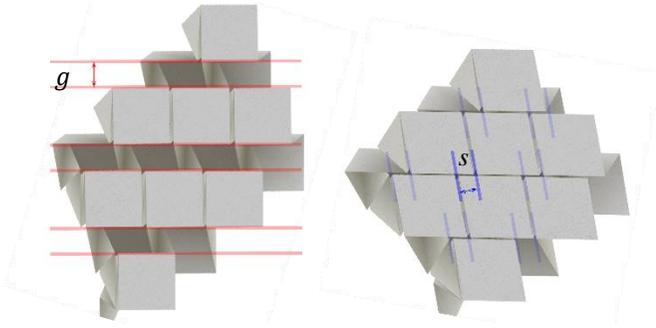

**Figure 11:** *Left: A gap distance g between skin-parallelogram columns of a final folded state. Right: A shift distance s of the phase of a double-tiling origami.*

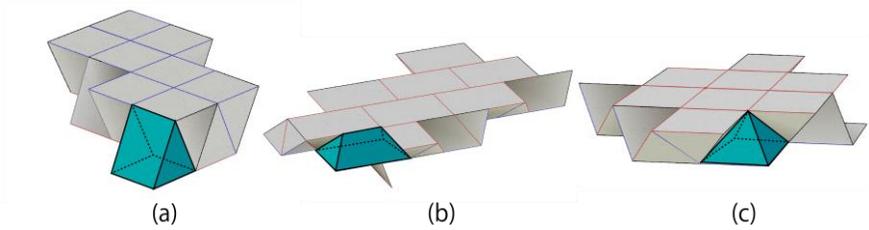

**Figure 12:** *Double-tiling origami patterns and their inner structures. (a) Prism type, (b) Hip-roof type, and (c) Pyramid type.*

| Shift distance $s$ | Inner structure | Name |
|---|---|---|
| $s = 0$ | Triangular prism, degenerated tetrahedron | Prism type |
| $0 < s < d$ | Hip-roof, tetrahedron | Hip-roof type |
| $s = d$ | Quadrangular cone, tetrahedron | Pyramid type, i.e., dual-tiling origami [Adachi et al. 2017] |

**Table 13:** *The relationship between inner structure and shift distance s.*

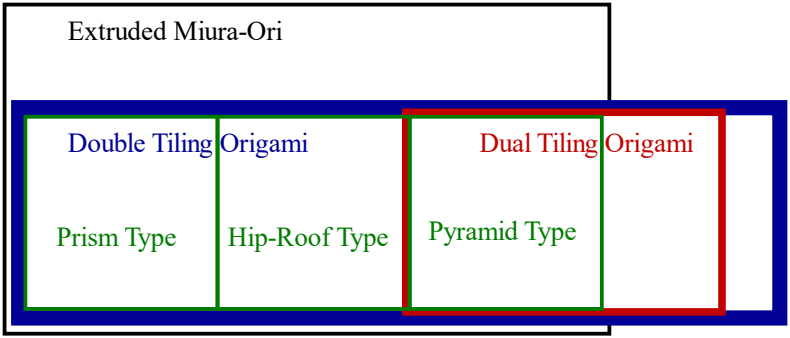

**Figure 14:** Conjectured inclusion relations between patterns of extruded Miura-Ori, double-tiling origami, and dual-tiling origami.

## 4 Discussion and Future Work

In this paper, we proposed a method of creating extruded Miura-Ori by splitting Miura-Ori at oblique cutting lines and extruding the edges to insert parallelogram strips between the split edges. Furthermore, we observed that there are three folded states whose skin-parallelograms lie on skin-planes. In the final folded state, within each column, skin-parallelograms share their edges. We can tweak the parameters to make an extruded Miura-Ori a double-tiling origami, where the skin-parallelograms tile the planes. Extruded Miura-Ori that conforms to double-tiling origami can be further classified by the types of polyhedra filling the space between planes.

As the concept of edge-extrusion of origami tessellation patterns is general, the approach can be applied to different origami tessellations to produce face-bondable origami tessellation patterns. The characterization of origami tessellations that allows the edge-extrusion approach is a topic for future studies, but we conjecture that like in Miura-Ori, iso-area symmetry is required to make the extrusion direction parallel to skin-planes.

## Appendix

**Dual-Tiling Origami**

One family of origami tessellation patterns that allow face-bonding to skin-planes is the dual-tiling origami family, proposed in [Adachi et al. 2017]. Parallelogram dual-tiling origami consists of parallelogram pyramids and tetrahedrons, and four parallelogram pyramids surround one tetrahedron (Figure 15). Looking at dual-tiling origami as a sandwich core, the bottom of each parallelogram pyramid tiles both planes. However, the dual-tiling origami core cannot ventilate because it consists of closed cells.

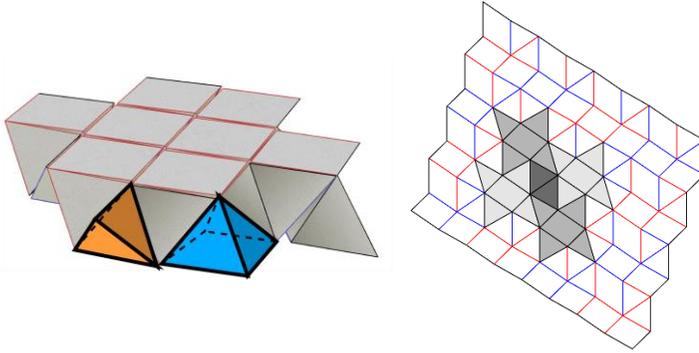

**Figure 15:** *A parallelogram dual-tiling origami and its crease pattern* [Adachi 2017].

The shape parameters (lengths of edges and the angle) of parallelograms on the skin-planes (as in the case of extruded Miura-Ori, we call these parallelograms *skin-parallelograms*) can be continuously changed. An interesting observation is that by continuously shortening the heights of the skin-parallelograms, the dual tiling origami degenerates into Miura-Ori by squashing parallelogram strips, shown in grey in Figure 16. The edge-extrusion approach is an analogy of this observation.

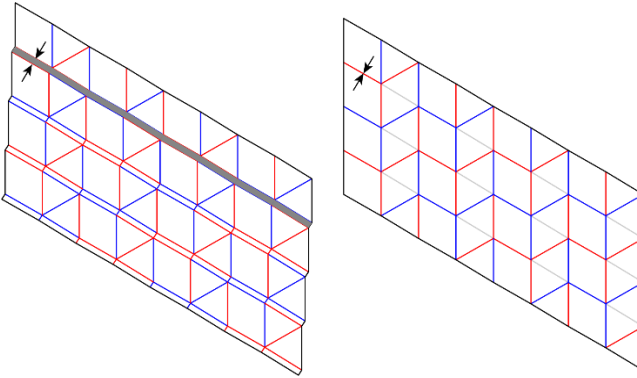

**Figure 16:** *Relationship between dual-tiling origami and Miura-Ori.*

| | (1) Development | (2) Geometric construction | (3) Final folded state |
|---|---|---|---|
| (a): The gap distance $G > 0$ | 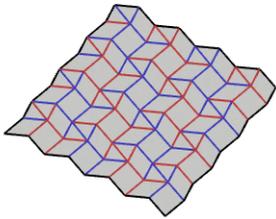 | 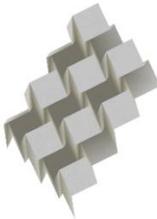 | 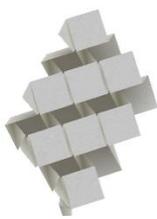 |
| (b): The gap distance $G > 0$ (1) and (2) merged | 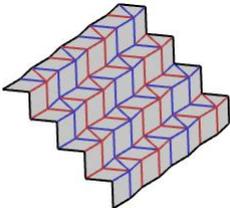 | | 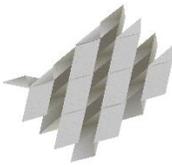 |
| (c): Prism Type $G = 0$ $S = 0$ (2) and (3) merged | 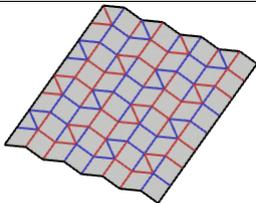 | 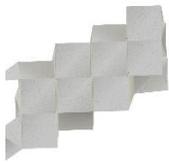 | |
| (d): Hip-Roof Type $G = 0$ $d > S > 0$ | 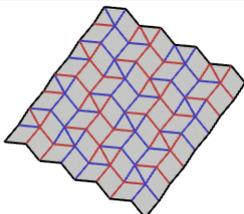 | 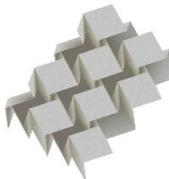 | 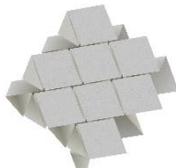 |
| (f): Pyramid Type (Dual Tiling Origami) $G = 0$ $S = d$ | 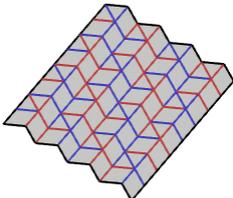 | 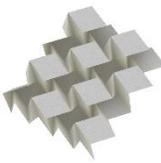 | 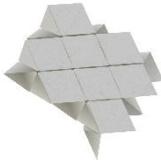 |

**Figure 17:** *Critical folded states of face-bondable origami tessellations. The folded states of (b) between the development and the geometrical construction are merged because we applied the edge-extrusion approach when the Miura-Ori was in the developed state. The folded states of (c) between the geometric construction and the final folded state are merged because we applied our approach when the Miura-Ori was in a flat folded state.*

Kai Suto
Graduate School of Arts and Sciences, the University of Tokyo, 3-8-1 Komaba, Meguro-ku, Tokyo 153-8902, Japan, e-mail: suto@graco.c.u-tokyo.ac.jp

Akito Adachi
Graduate School of Arts and Sciences, the University of Tokyo, 3-8-1 Komaba, Meguro-ku, Tokyo 153-8902, Japan, e-mail: adachi.akito.53u@gmail.com

Tomohiro Tachi
Graduate School of Arts and Sciences, the University of Tokyo, 3-8-1 Komaba, Meguro-ku, Tokyo 153-8902, Japan, e-mail: tachi@idea.c.u-tokyo.ac.jp

Yasushi Yamaguchi
Graduate School of Arts and Sciences, the University of Tokyo, 3-8-1 Komaba, Meguro-ku, Tokyo 153-8902, Japan, e-mail: yama@graco.c.u-tokyo.ac.jp